\documentclass[aps,prl,reprint,superscriptaddress]{revtex4-1}
\usepackage{graphicx}  
\usepackage{dcolumn}   
\usepackage{bm}        
\usepackage{amssymb}   
\usepackage{epstopdf}

\usepackage[caption=false]{subfig}
\usepackage[colorlinks=true, breaklinks=true, bookmarks=false, linkcolor=blue, urlcolor=blue, citecolor=blue,]{hyperref}

\begin{document}

\title{Weak Localization Effects as Evidence for Bulk Quantization in Thin Films Bi$_2$Se$_3$}

\author{Li Zhang}
\affiliation{Department of Applied Physics, Stanford University, Stanford, CA 94305, USA}
\affiliation{Geballe Laboratory for Advanced Materials, Stanford University, Stanford, CA 94305, USA}

\author{Merav Dolev}
\affiliation{Department of Applied Physics, Stanford University, Stanford, CA 94305, USA}
\affiliation{Geballe Laboratory for Advanced Materials, Stanford University, Stanford, CA 94305, USA}

\author{Qi I. Yang}
\affiliation{Department of Physics, Stanford University, Stanford, CA 94305, USA}
\affiliation{Geballe Laboratory for Advanced Materials, Stanford University, Stanford, CA 94305, USA}

\author{Robert H. Hammond}
\affiliation{Geballe Laboratory for Advanced Materials, Stanford University, Stanford, CA 94305, USA}

\author{Bo Zhou}
\affiliation{Department of Physics, Stanford University, Stanford, CA 94305, USA}
\affiliation{Department of Physics, University of Oxford, Parks Road, Oxford OX1 3PU, United Kingdom}
\affiliation{Advanced Light Source, Lawrence Berkeley National Laboratory, Berkeley, California 94720, USA}

\author{Alexander Palevski}
\affiliation{School of Physics and Astronomy, Tel Aviv University, 69978 Tel Aviv, Israel}

\author{Yulin Chen}
\affiliation{Department of Physics, Stanford University, Stanford, CA 94305, USA}
\affiliation{Department of Physics, University of Oxford, Parks Road, Oxford OX1 3PU, United Kingdom}
\affiliation{Advanced Light Source, Lawrence Berkeley National Laboratory, Berkeley, California 94720, USA}

\author{Aharon Kapitulnik}
\affiliation{Department of Applied Physics, Stanford University, Stanford, CA 94305, USA}
\affiliation{Geballe Laboratory for Advanced Materials, Stanford University, Stanford, CA 94305, USA}
\affiliation{Department of Physics, Stanford University, Stanford, CA 94305, USA}


\begin{abstract}
Strong spin-orbit coupling in topological insulators results in the ubiquitously observed weak antilocalization feature in their magnetoresistance. Here we present magnetoresistance measurements in ultra thin films of the topological insulator Bi$_2$Se$_3$, and show that in the 2D quantum limit, in which the topological insulator bulk becomes quantized, an additional negative magnetoresistance feature appears. Detailed analysis associates this feature with weak localization of the quantized bulk channels, providing thus evidence for this quantization. Examination of the dephasing fields at different temperatures indicates different scattering mechanism in the bulk vs the surface states. 
\end{abstract}

\maketitle

A three-dimensions (3D) topological insulator (TI) is fully gapped in the bulk, but exhibits an odd number of surface 2D massless cones of helical Dirac fermions, which are protected by time reversal symmetry (TRS) and thus cannot be destroyed by any non-magnetic impurities \cite{Kane_Rev, Zhang_Rev}.  The Dirac fermions are helical in the sense that the electron spin points perpendicularly to the momentum, forming a lefthanded helical texture in momentum space. This strong coupling of spin and momentum leads to a range of new phenomena, especially when the TI is brought in contact with either a superconductor or an insulating ferromagnet, giving rise to possible observations of Majorana fermions \cite{Fu_Kane_Majorana,Moore_Rev}, the topological magneto-electric effect \cite{Zhang_magnetoelectric}, and quantum anomalous Hall effect \cite{Zhang_QAHE}.

An important  consequence of spin-momentum locking is the full suppression of backscattering resulting in a relative $\pi$ Berry phase acquired by electrons executing time-reversed paths.  As a consequence, at low temperatures, when the dephasing length ($\ell_\phi$) of the surface state electrons is long, this results  in destructive interference, which give rise to positive quantum corrections $\Delta \sigma  > 0$, to the Drude conductivity. The result of these quantum corrections to the conductivity are called weak antilocalization (WAL) effects, and are expected to occur in general when spin-orbit (SO) interaction is strong \cite{WAL_theory, WAL_Hikami, Culcer_Rev}. When weak magnetic field $H\gtrsim H_\phi \equiv \Phi_0/(8\pi \ell_\phi^2)$ ($\Phi_0=hc/e$ is a flux quantum) is applied, these interference effects are reduced giving rise to positive magnetoresistance (MR) which is therefore a hallmark of WAL.

While an ideal TI is a true bulk insulator, this property has proven to be very difficult to achieve. Interstitials, vacancies, and antisite doping, are only  a few of the common issues that give rise to a substantial bulk carrier density causing the chemical potential to be pinned at the bulk conduction band. Focusing on the Bi$_2$Se$_n$Te$_{3-n}$ system (where n=0,1,2 and 3), partial mitigation of the problem is commonly achieved by compensation (e.g. doping  Bi$_2$Se$_3$ with Sb) \cite{Pb_doping,Analytis2010}, or by controlling the Fermi energy via a gate bias \cite{Chen_BG1, Chen_BG2, Hadar_top_gate}.  However, unless the Fermi energy is tuned to be much below the bottom of the bulk conduction band, transport in such TI systems has been shown to be a complicated combination of surface and bulk states, which in principle is not a simple sum of parallel conduction paths due to surface-bulk interactions \cite{WL_Glazman}. Furthermore, as the 3D TI becomes thin enough to reach the 2D quantum limit, the 3D bulk becomes quantized, and the bulk transport is through the quantized 2D channels, which may contribute negative MR for low enough Fermi energy \cite{WL_bulk_Lu, WL_Glazman}.

MR measurements on thin film TIs ubiquitously show a positive MR feature associated with WAL \cite{ Chen_BG1, Chen_BG2, Rutgers_thickness, Hadar_top_gate, TI_interaction, Pb_doping, He_BiTe, Li_MBE, Thick_ind, Ando_SdH}. An additional negative MR feature was found in magnetically doped TI films \cite{Cr_doped, 2012Beidenkopf}, and in a few cases, also in undoped TI thin films \cite{Ando_SdH, WL_WAL_gated}. In the case of short elastic mean free path, no spin-flip events, and zero or infinite SO interaction strength, the magnetoconductance follows the expression \cite{WAL_Hikami}:
\begin{equation}
 \Delta \sigma (H) \equiv \sigma(H)-\sigma(0)=  \alpha \sigma_0 f \left(  \frac{H_\phi}{H} \right) ,
 \label{eq:simple_WAL} 
\end{equation} 
where $\sigma_0 \equiv e^2/(2 \pi^2 \hbar)$, $H_\phi$ the characteristic  dephasing field, $f(z)\equiv ln(z) -\psi (1/2+z)$, in which $\psi$ is the digamma function, $H$ the external applied magnetic field, and $\alpha=-1$  ($\alpha=1/2$) for zero (infinite) SO interaction strength. The commonly used way of analyzing MR data in TIs is to fit it to eq.~\ref{eq:simple_WAL}, and use the prefactor $\alpha$ as a fitting parameter. This procedure usually yields $0.3<\alpha<1$, which is interpreted as an indication for contribution from one or more surface states. 

However, this type of analysis is not in line with theory. For finite SO interaction, eq.~\ref{eq:simple_WAL} is over simplified, and the full expression by Hikami {\it et al.} \cite{WAL_Hikami}, which includes a finite SO interaction term, should be used, allowing the magnitude of the SO interaction term to vary, while keeping the prefactor of the function $f$ fixed. For ultrathin TI films, there is a need to include in the analysis not only the surface states with infinite SO interaction that leads to WAL, but also the contribution of the quantized $2$D bulk band channels, and their possible WL contribution \cite{WL_bulk_Lu, WL_Glazman}. 

In this paper we present MR measurement results taken on three different ultrathin Bi$_2$Se$_3$ samples, and show that on top of the commonly observed positive MR associated with WAL, an additional negative MR feature, associated with WL, appears in those films at high magnetic fields. Careful analysis of the MR reveals contribution of WAL from the two surface states and of WL from two quantized bulk band channels, suggesting direct observation of bulk quantization in ultrathin films of Bi$_2$Se$_3$.

\begin{figure}[h]
\centering
\subfloat[]{\includegraphics[width=0.55\linewidth]{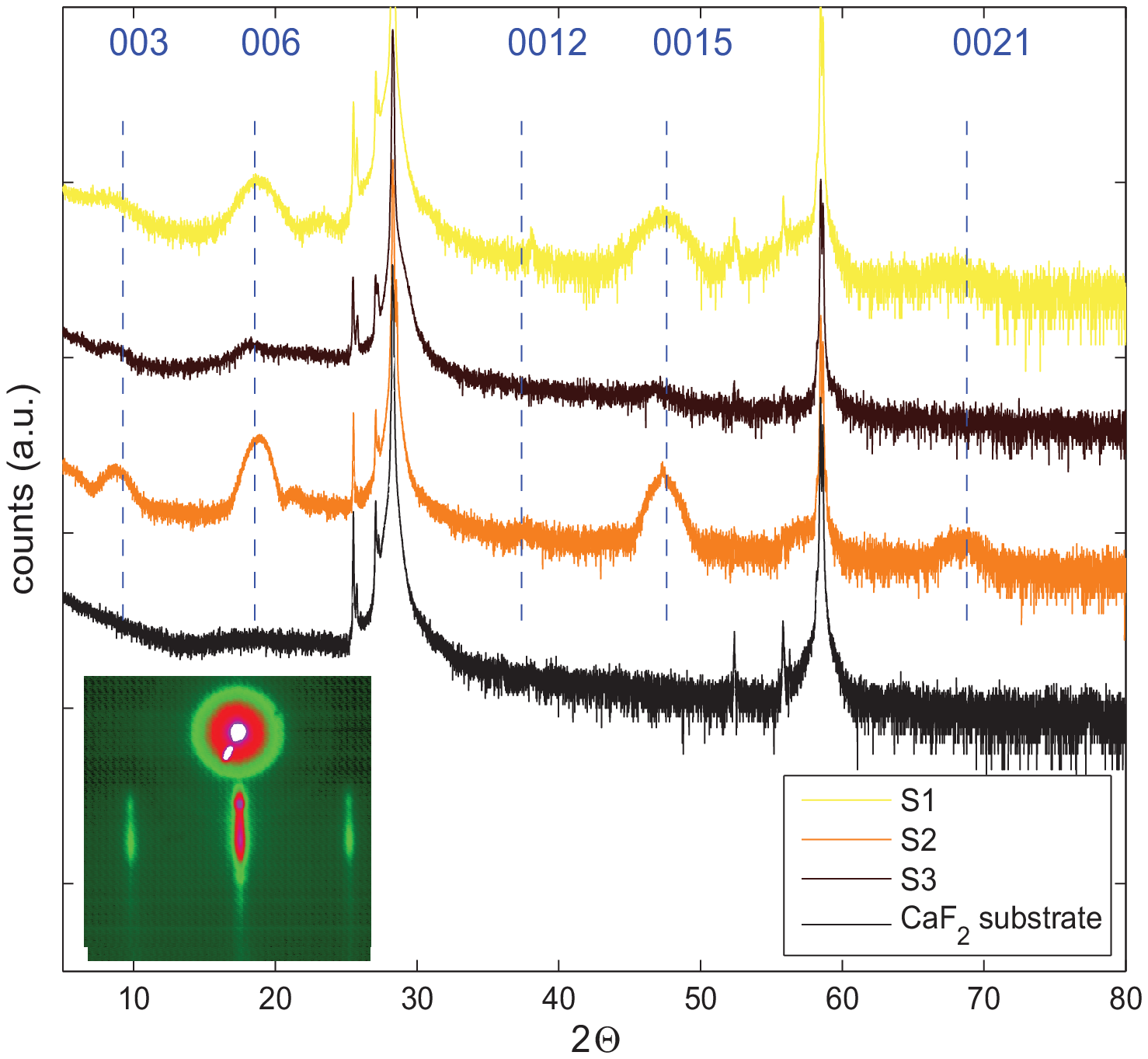}
\label{fig:XRD_RHEED}}
\subfloat[]{\includegraphics[width=0.4\linewidth]{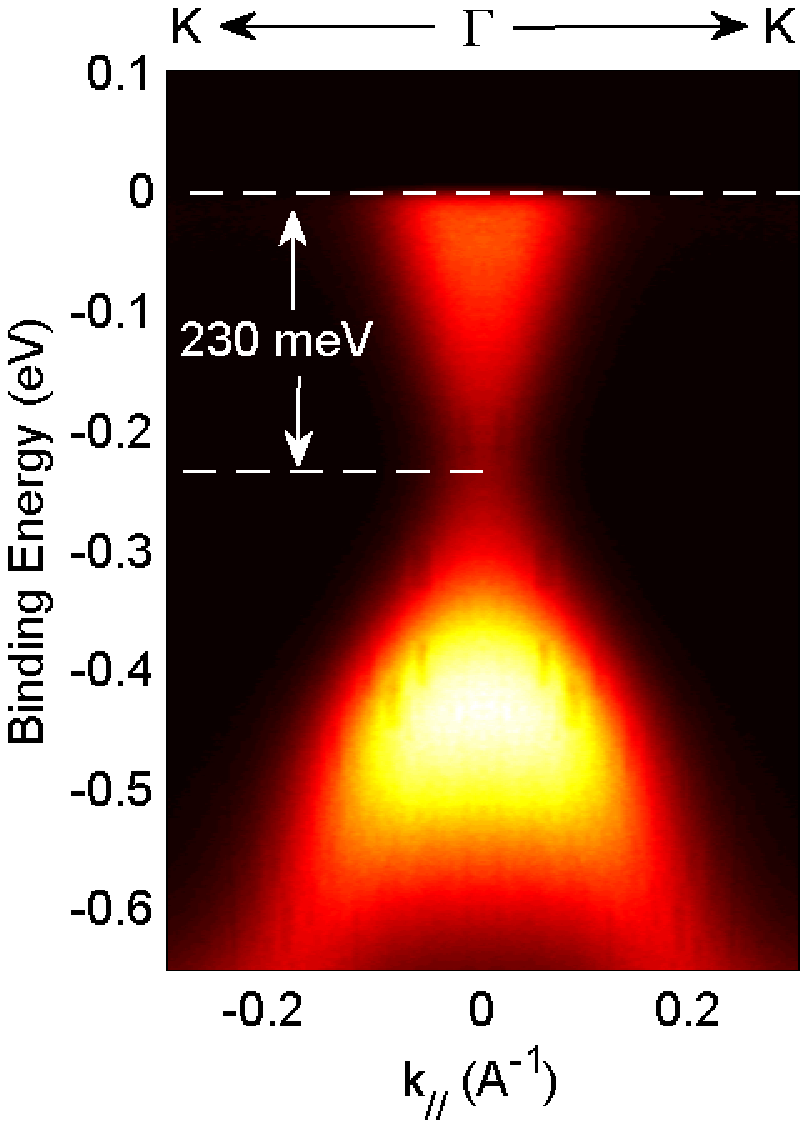}\label{fig:ARPES}}
\\
\caption{(Color online) \protect\subref{fig:XRD_RHEED}~X-ray Diffraction data on all samples showed clear c-oriented Bi$_2$Se$_3$ peaks. The streaky RHEED pattern of S$1$ (inset on the left bottom) taken \textit{in situ} indicates good crystallinity and smooth surface.\protect\subref{fig:ARPES}~ ARPES of S$4$, showing a relatively low Fermi energy above the Dirac point.}
\label{fig:growth}
\end{figure}

\begin{figure}[h]
\centering
{\includegraphics[width=\linewidth]{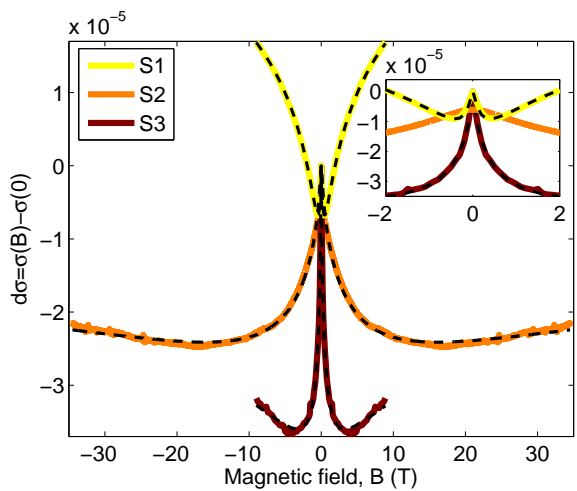}\label{fig:WL_WAL_2K}}
\caption{ (Color online) Magnetoresistance of samples S$1$, S$2$, S$3$, measured at 2K. Colorful lines (yellow/orange/brow) are the measured data, while the black dashed lines are the fittings, which assume two surface states (both with infinite SO interaction) and two bulk bands (both with zero or small effective SO interaction). The inset shows the same data and fit curves at low magnetic fields. The fitting parameters are given in table \ref{tab:FitCoeff_2K}.}
\label{fig:MR_2K}
\end{figure}

The ultrathin Bi$_2$Se$_3$ samples studied were grown on CaF$_2$ substrates using molecular beam epitaxy (MBE), as described in \cite{Li_MBE}. Three samples, labelled S$1$, S$2$, S$3$ with films thickness of $4$ nm, $5$ nm, $7$ nm respectively, were patterned and measured in a Van der Pauw (VdP) configuration. A fourth sample S$4$, $6$ nm thick, was grown immediately after sample S$2$, with identical growth conditions, and examined by angle resolved photoemission spectroscopy(ARPES). High crystallinity of the ultrathin films were confirmed by both the \textit{in situ} reflected high energy electron diffraction (RHEED) pattern and \textit{ex situ} X-ray Diffraction data, as shown in Fig.~\ref{fig:XRD_RHEED}. ARPES on sample S$4$, presented in Fig.~\ref{fig:ARPES}, shows a low Fermi energy level.  Therefore, thin samples in which the bulk band is quantized should have only a few full $2$D quantized bulk channels that can contribute to the conductance (assuming quantization and gap opening similar to the one observed via ARPES in other high quality ultrathin Bi$_2$Se$_3$ films \cite{Zhang_ARPES_thin}).
Magnetoresistance measurements (sheet resistance as a function of magnetic field) were obtained at $2K$ on samples S$1$, S$2$, S$3$, with magnetic fields ranging between $-9$ T and $9$ T for samples S$1$ and S$3$, and between $-35$ T and $35$ T for sample S$2$, as shown in Fig.~\ref{fig:MR_2K}. All samples show negative MR at high fields, which we interpret as WL contribution of the quantized bulk band channels. The strongest WL contribution was observed in sample S1, which is the thinnest ($4$ nm) Bi$_2$Se$_3$ film that was measured. However, the magnitude of the WL and the magnetic field at which the MR switches from positive to negative depends not only on the sample thickness, but also on the sample quality, and the corresponding dephasing lengths for the different surface and bulk bands. For example, the negative MR in the $5$ nm thick sample S$2$ was observed in a much higher magnetic field, compared to the $7$ nm sample S$3$. 
We note that S$2$ has an additional low field feature, due to which we shifted the MR of S$2$ accordingly for the WAL and WL analysis. This low field feature will be discussed in a separate publication \cite{TBD}. 

The measured magnetoconductance $\Delta \sigma (H) \equiv \sigma(H) - \sigma(0)$ was fitted to the full expression by Hikami {\it et al.} \cite{WAL_Hikami} for independent conduction channels: 
\begin{equation}
\Delta \sigma (H) =  \sigma_0 \sum_\alpha \left[f \left( \frac{H^\alpha_1}{H} \right) -\frac{3}{2} f \left( \frac{H^\alpha_2}{H} \right) +\frac{1}{2} f \left( \frac{H^\alpha_3}{H} \right) \right],
\label{eq:WAL_ind}
\end{equation}
where the summation of $\alpha$ takes into account all conductance channels, and
\begin{equation}
\begin{array}{l}
H^\alpha_1=H_e+H^\alpha_{SO}+H_s\\
\\
H^\alpha_2=\frac{4}{3}H^\alpha_{SO}+\frac{2}{3}H_s+H^\alpha_\phi\\
\\
H^\alpha_3=2H_s+H^\alpha_\phi .
\end{array}
\end{equation}
The characteristic fields $H_n$ are related to the characteristic lengths $\ell_n$ and times $\tau_n$ by $H_n= \hbar /(4e \ell^2_n)= \hbar /(4e D \tau_n)$, where $H_e$, $H_s$, $H^\alpha_{SO}$, $H^\alpha_{\phi}$ are the characteristic fields for elastic scattering, spin-flip events, SO interaction in band $\alpha$, and 
dephasing in band $\alpha$ respectively.

In our fitting we assumed no spin-flip events, namely, $H_s=0$, and an elastic scattering length $\ell_e<8$nm, resulting in the characteristic elastic field being $H_e > 15T$ and thus in a negligible $f \left( \frac{H^\alpha_1}{H} \right)$ (the fitting is hardly affected by a change of $H_e$, for any $H_e > 15T$). The best fit was obtained when taking into account two surface states (labeled as $\alpha=s1,s2$) and two quantized bulk channels (labeled as $\alpha=b1,b2$). For the two surface states, the SO interaction was set to be $H^{s1,s2}_{SO}=\infty$, as predicted for TI surface states. However, despite the fact that the SO interaction in TIs is very strong, quantized bulk band channels with small Fermi surface (the Fermi energy is low with respect to the bottom of the conduction band) will have an effective small SO interaction (and a corresponding small $H^{b}_{SO}$) \cite{WL_bulk_Lu, WL_Glazman}. In agreement with the low Fermi energy seen in ARPES (Fig.~\ref{fig:ARPES}), $H^{b1,b2}_{SO}$ of the quantized bulk channels was found in our analysis to be zero or finite and small, leading to a WL contribution of the bulk channels. The characteristic SO and dephasing 
fields for the different conduction channels, as calculated from our $2K$ MR data, are given in table \ref{tab:FitCoeff_2K}, with the corresponding fits plotted in dashed in Fig.~\ref{fig:MR_2K}. 

\begin{table}
	\centering
		\begin{tabular}{|l|l|l|l|l|l|l|l|}
			          \hline Sample & $H^{s1,s2}_{SO}$ & $H^{b2}_{SO}$ & $H^{b1}_{SO}$ & $H^{s1}_\phi$ & $H^{s2}_\phi$ &   $H^{b1}_\phi$ & $H^{b2}_\phi$\\ 
			          \hline
							 	 $S1$   & $\infty$ & $0.1$ & $0.08$ & $0.224$ &   $0.647$  &  $0.017$  &  $0.0498$ \\
	          		 $S2$   & $\infty$ & $0.05$ & $0$ & $ 0.661$  &  $0.034$  &  $2.83$  &  $2.93$ \\
	          		 $S3$   & $\infty$ & $0.2$ & $0$ & $ 0.124 $ & $   0.0297 $ & $   0.0037 $ & $   0.83$ \\    \hline         
		\end{tabular}
	\caption{The characteristic fields associated with SO interaction ($H^{\alpha}_{SO}$) and dephasing 
	($H^{\alpha}_{\phi}$), for the two surface states marked as $\alpha=s1,s2$ and for the two quantized bulk channels marked as $\alpha=b1,b2$, as found from fitting the $2K$ MR data (Fig.~\ref{fig:MR_2K}), for all three samples.}
	\label{tab:FitCoeff_2K}
\end{table}

\begin{figure}[h]
\centering
{\includegraphics[width=\linewidth]{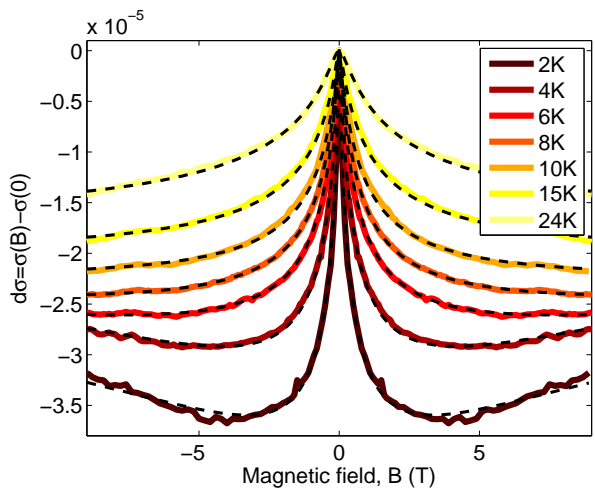}\label{fig:WL_WAL_43M}}
\caption{(Color online) Magnetoresistance of sample S3, measured at temperatures ranging from $2K$ to $24K$. Data is presented together with our fitting curve (in dashed black), which assumes two surface states (both with infinite SO interaction) and two bulk bands (both with zero or small effective SO interaction). The fitting parameters are given in table \ref{tab:FitCoeff_43M}.}
\label{fig:MR_43M}
\end{figure}

\begin{figure}[h]
\centering
{\includegraphics[width=\linewidth]{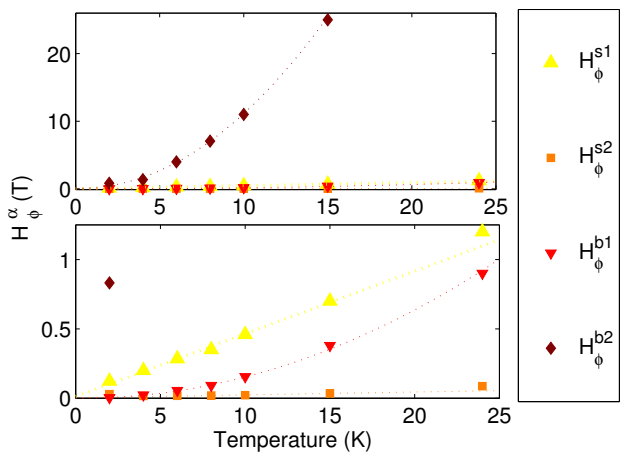}}
\caption{(Color online) 
Dephasing fields as a function of temperature, calculated from fitting the MR of sample S3. The upper plot shows the full $H_\phi$ range, up to $26T$, while the lower plot zooms into the low $H_\phi$ range. The dashed lines are guidance to the eye, showing the linear temperature dependence of the surface states' $H_\phi$'s and quadratic temperature dependence of the bulk channels' $H_\phi$'s on the temperature.}
\label{fig:Hi_T}
\end{figure}

To further study the high-field WL feature, we measured the MR of sample $S3$ at different temperatures, as shown in Fig.~\ref{fig:MR_43M}.  
Fitting the measured magnetoconductance with the expression for independent conduction channels given in eq.~\ref{eq:WAL_ind}, we find the characteristic 
dephasing fields $H^\alpha_\phi$ at each temperature, given in table \ref{tab:FitCoeff_43M}. Examination of the temperature dependence of these dephasing fields, presented in Fig.~\ref{fig:Hi_T}, shows a linear increase of the surface dephasing fields $H^{s1,s2}_\phi$  with temperature, as expected for electron-electron scattering at low temperatures \cite{AAK_Hi_T}:
\begin{equation}
\tau^{-1}_\phi= \frac{k_B e^2}{2 \pi \hbar^2} R_{\Box}{\rm ln}[\sigma] T ,
\end{equation}
where $R_\Box$ is the sheet resistance of the film, and $\sigma$ is the dimensionless conductance of the sample. However, for the quantized bulk channels, we find  $H^{b1,b2}_\phi \propto aT^2 + bT$.  Such a behavior can be a consequence of two different dephasing mechanisms originating from electron-electron interaction (linear term) and electron-phonon interaction (quadratic term) which add up for an effective inverse scattering time\cite{Lin_Rev_dephasing}:
\begin{equation}
\tau_\phi^{-1}=\tau_{e-e}^{-1}+\tau_{e-ph}^{-1},
\end{equation}
Alternatively, it can be a consequence of pure electron-electron interaction in an intermediate regime of temperatures in which the quadratic term (dominant at high temperatures) is still significant. In this case, the fact that the quadratic term still appears in the bulk bands while the surface bands show only a linear temperature dependence suggests  different temperature threshold at which large energy transfer scattering processes become dominant, for bulk and surface states.  Regardless of the mechanism which is responsible for the 
dephasing in the bulk bands, it is clear that there is a distinguishable difference between the dephasing 
mechanisms of the surface states and the quantized bulk channels.

While we obtain excellent fits to our data in a wide range of temperatures and magnetic fields, it is important to note that we are using a rather simple model, which does not include interaction between different conduction channels, either bulk or surface.  A model that includes interactions in a limited form was recently introduced by Garate and Glazman \cite{WL_Glazman}, who derived approximate analytical expressions for the low-field magnetoconductance of one bulk and one surface band, including interactions among these two channels.  Their results reveal parameter regimes for both WL and WAL, which in principle is in line of our observations. However, our attempt to use this model failed, possibly indicating that more than just two bands are present. In addition, we found that spin-orbit interaction should be treated as intermediate in some of the bulk bands. 
Since we believe that interactions among the different conduction channels have to be present, it is possible that further development of this model is needed to allow better fitting to our data.

\begin{table}
	\centering
		\begin{tabular}{|l|l|l|l|l|l|l|l|}
			      \hline $T$ & $H^{s1,s2}_{SO}$ & $H^{b2}_{SO}$ & $H^{b1}_{SO}$ & $H^{s1}_\phi$ & $H^{s2}_\phi$ &   $H^{b1}_\phi$ & $H^{b2}_\phi$\\ \hline					
							 	 $2K$   & $\infty$ & $0.2$ & $0$ & $ 0.124 $ & $   0.0297 $ & $   0.0037 $ & $   0.83$ \\
	          		 $4K$   & $\infty$ & $0.2$ & $0$ & $ 0.2 $ & $   0.0158 $ & $   0.0225 $ & $   1.35$ \\
	          		 $6K$   & $\infty$ & $0.2$ & $0$ & $ 0.285 $ & $   0.0172 $ & $   0.055 $ & $   4$ \\
							 	 $8K$   & $\infty$ & $0.2$ & $0$ & $ 0.351 $ & $   0.0188 $ & $   0.0911 $ & $   7$ \\ 
							 	 $10K$   & $\infty$ & $0.2$ & $0$ & $ 0.46 $ & $   0.022 $ & $   0.155 $ & $  11$ \\ 
							 	 $15K$   & $\infty$ & $0.2$ & $0$ & $0.7  $ & $  0.034  $ & $  0.38  $ & $ 25$ \\ 
							 	 $24K$   & $\infty$ & $0.2$ & $0$ & $1.2  $ & $  0.0871 $ & $   0.9 $ & $  66$ \\ \hline 
		\end{tabular}
	\caption{The characteristic SO and dephasing 
	fields for sample S3, as found from fitting the MR data taken at different temperatures (figure \ref{fig:MR_43M}).}
	\label{tab:FitCoeff_43M}
\end{table}

In summary, magnetoresistance measurements of ultrathin  films of the topological insulator Bi$_2$Se$_3$ reveal, in addition to the ubiquitous positive magnetoresistance typically observed in such films and attributed to WAL effects in the surface state band, also a negative magnetoresistance feature at high magnetic fields.  Using  ideas first introduced by Lu and Shen \cite{WL_bulk_Lu}, we show excellent fits of our data to a simple model of two surface states and two quantized bulk bands. Our work therefore suggests a direct observation of bulk band quantization in thin films of Bi$_2$Se$_3$.

\begin{acknowledgments}
The authors would like to thank Nick Breznay, Boris Spivak, Leonid Glazman and Ion Garate for fruitful discussions. A portion of this work was performed at the National High Magnetic Field Laboratory, which is supported by National Science Foundation Cooperative Agreement No. DMR-1157490, the State of Florida, and the U.S. Department of Energy. The authors would also like to thank Alexey V. Suslov and Scott A. Maier for all their time and help at NHMFL. This research project is supported by DARPA MESO project (No. N66001-11-1-4105), and by FENA, and by a seed grant from the Department of Energy for the study of topological insulators.
\end{acknowledgments}

\bibliography{WLWAL}

\end{document}